\documentclass[twocolumn,a4paper,amsmath,amssymb]{revtex4}
\usepackage{graphicx}
\usepackage{amsmath}
\usepackage{braket}
\usepackage{xcolor}
\usepackage{natbib}
\usepackage{dsfont}

\renewcommand{\eqref}[1]{Eq.~(\ref{#1})}

\newcommand{\AW}[1]{{\color{black}{#1}}}

\newcommand{\removedD}[1]{{\color{gray}{#1}}}
\renewcommand{\removedD}[1]{{}} 
\newcommand{\CE}[1]{{\color{black}{#1}}}

\begin{document}
\title{Observation of Entanglement Between Itinerant Microwave Photons and a Superconducting Qubit}
\author{C.~Eichler, C.~Lang, J.~M.~Fink, J.~Govenius, S.~Filipp, and A.~Wallraff}
\affiliation{Department of Physics, ETH Z\"urich, CH-8093, Z\"urich, Switzerland.}
\date{\today}
\begin{abstract}
{A localized qubit entangled with a propagating quantum field is well suited to study non-local aspects of quantum mechanics and may also provide a channel to communicate between spatially separated nodes in a quantum network. Here, we report the on demand generation and characterization of Bell-type entangled states between a superconducting qubit and propagating microwave fields composed of zero, one and two-photon Fock states. Using low noise linear amplification and efficient data acquisition we extract all relevant correlations between the qubit and the photon states and demonstrate entanglement with high fidelity.}
\end{abstract}
\maketitle
One of the most fascinating aspects of quantum physics is the entanglement between two spatially separated objects sharing a common non-local wave function. Propagating photons are ideal carriers for distributing such entanglement between distant matter systems in a quantum network.
Entanglement between  photons and stationary qubits has so far been exclusively studied at optical frequencies with single atoms~\cite{Blinov2004,Volz2006,Stute2012} and electron spins~\cite{Togan2010}, to interface stationary and flying qubits~\cite{Wilk2007}, to implement quantum teleportation~\cite{Olmschenk2009,Moehring2004} and to realize nodes for quantum repeaters~\cite{Yuan2008} and networks~\cite{Ritter2012,Kimble2008,Moehring2007}. Rapid progress in the development of superconducting circuit based quantum technologies also renders propagating \cite{Houck2007,Bozyigit2011,Mallet2011,Eichler2011,Flurin2012} and localized microwave photons~\cite{Haroche2006,Hofheinz2009} an attractive carrier of quantum information. A major obstacle in measuring quantum correlations between superconducting artificial atoms and itinerant photons has so far been the limited detection efficiency at microwave frequencies. Here, we overcome this problem by using a quantum limited amplifier \cite{Castellanos2008}, which significantly improves the signal-to-noise ratio in both photon field and qubit measurement.  In combination with novel tomography methods \cite{Eichler2012} this allows us to measure quantum correlations between itinerant microwave radiation and a stationary qubit with high fidelity.

In our experiments we create entangled states between a superconducting transmon qubit~\cite{Koch2007} and an itinerant microwave radiation field containing up to two photons. The field mode $a$ can be described by two canonically conjugate variables ${X}$ and ${P}$ analogous to the position and momentum variables of a mechanical quantum harmonic oscillator. In contrast to most experiments performed at optical frequencies, we simultaneously measure both continuous variables $X,\,P$ rather than the photon number of the field. This \AW{enables} us to fully characterize quantum fields \AW{beyond the single photon level}. In addition to the measurement of photon statistics of the field, we  determine the correlations between the measured qubit state and the observed values $X$ and $P$ which clearly demonstrate that the qubit is entangled with the quantum field.

We deterministically prepare $1.25 \times 10^5$ Bell states of the form $|\psi\rangle=(|0e\rangle+|1g\rangle)/\sqrt{2}$ per second, in which a single excitation is shared coherently between a qubit and a single propagating mode of a radiation field. Here, $|g\rangle,|e\rangle$ label the qubit basis states and $|0\rangle,|1\rangle,|2\rangle,...$ the photon number states. To entangle the qubit and the radiation field we first bring the qubit from the ground state \CE{$|0g\rangle$} to the excited state \CE{$|0e\rangle$} by applying a \AW{$10 \, \rm{ns}$ long} $\pi$-pulse resonant with its transition frequency $\omega_{ge}/2\pi = 6.442 \,{\rm GHz}$. \AW{By applying a magnetic flux pulse we then tune the qubit into resonance with a transmission line resonator at frequency $\omega_{\rm r}/2\pi=7.133  \,{\rm GHz}$, which is strongly coupled to the qubit with rate $g/2\pi = 65 \,{\rm MHz}$~\cite{Wallraff2004}. After an interaction time of $\tau=\pi/4g \approx 2 \,{\rm ns}$ we obtain the state $(|0e\rangle+|1g\rangle)/\sqrt{2}$} \AW{up to a  phase factor which is omitted here for convenience}.

\begin{figure}[b]
\centering
\includegraphics[scale=1]{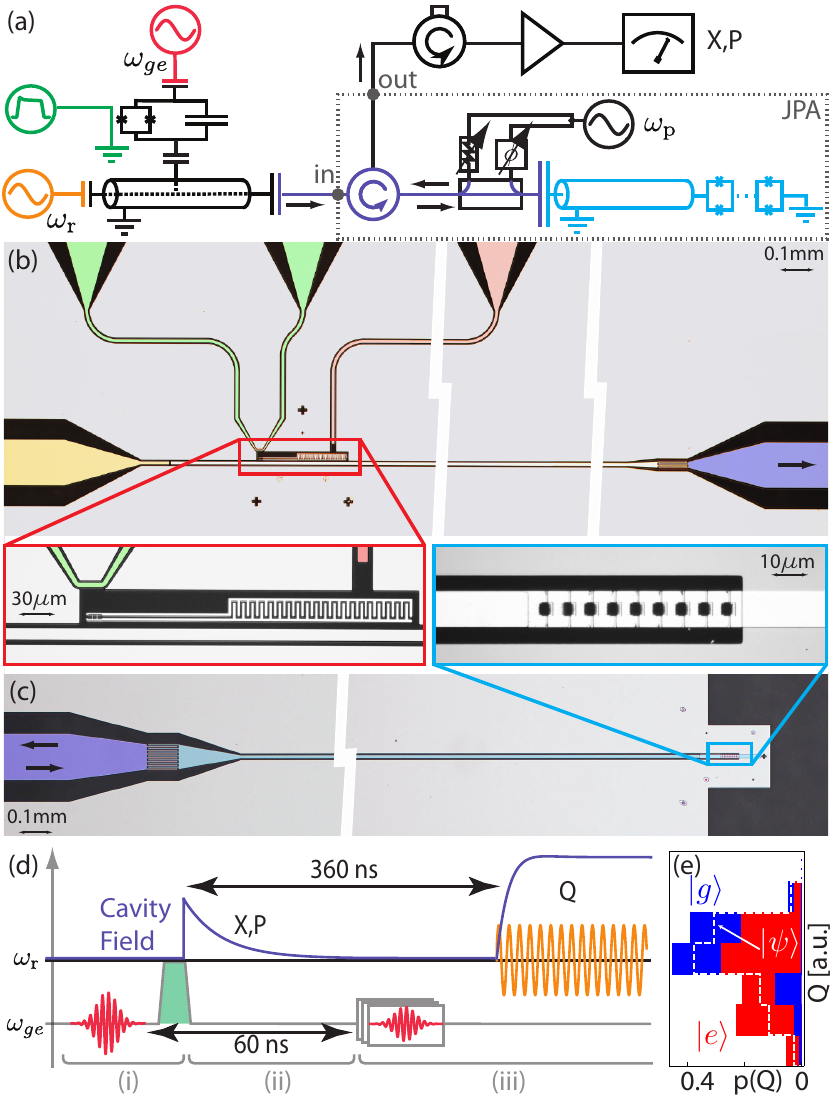}
\caption{{
Schematic of the experimental setup}. ({a}) A transmon qubit with individual charge drive line (red) at frequency $\omega_{ge}$ and flux control line (green) is strongly coupled to a resonator with a weakly coupled  input port \CE{($\gamma_{\rm in}/\kappa \approx0.001$)}  driven at frequency $\omega_{\rm r}$ for qubit readout (orange). The output port is strongly coupled into a transmission line (violet) and amplified with a Josephson parametric amplifier (light blue) pumped at $\omega_p$ through a directional coupler with adjustable phase and attenuation for pump tone cancelation. The signal reflected off the parametric amplifier passes through a chain of circulators into a low-noise semiconductor-amplifier after which its two quadratures $X, P$ are detected.  ({b}) False color optical micrograph of the sample. The transmon qubit (enlarged) consists of two capacitively coupled islands connected by a pair of Josephson junctions. ({c}) False color micrograph of the Josephson parametric amplifier. The array of Josephson junctions (enlarged) at the end of the quarter wavelength resonator (light blue) provides the nonlinearity. ({d}) Pulse sequence used for the experiment: (i) state preparation, (ii) field measurement and (iii) qubit read-out (see text for details). ({e}) Measured probability distribution $p(Q)$ of the qubit read-out quadrature $Q$ for prepared ground (blue), excited  (red) and Bell \CE{(dashed white)} states.
}
\label{fig:setup}
\end{figure}

The setup employed for this experiment is shown schematically in Fig.~\ref{fig:setup}(a) with a  micrograph of the sample shown in B and the experimental sequence in D. While the qubit on average keeps its excitation during its life time $T_1=1.0 \,\mu s$, the resonator field is emitted during the much shorter cavity decay time of $1/\kappa= 25 \,{\rm ns}$. Note that, the interaction time $\tau$ during state preparation is small compared to $1/\kappa$, which itself is small compared to the qubit life and coherence times ($T_2^*=220 \,ns$). This hierarchy of timescales ($1/g < 1/\kappa<T_1,T_2^*$) guarantees that the entangled state can be coherently prepared, and that the qubit remains in the excited state while the photon is emitted into the propagating transmission line mode $a$ of which both conjugate field quadratures $X$ and $P$ are detected (Fig.~\ref{fig:setup}(d)).

We measure the quadratures $X,\,P$ using a parametric amplifier operating close to the quantum limit~\cite{Castellanos2008}. The amplifier is based on a quarter wave transmission line resonator shunted by an array of Josephson junctions providing the Kerr nonlinearity used in the parametric amplification process~\cite{Yurke2006}, see Fig.~\ref{fig:setup}(c). We operate the parametric amplifier in a phase-preserving mode in which both conjugate field quadratures are amplified equally \cite{Eichler2011a}. This is achieved by pumping the amplifier at a frequency $12.5 \,{\rm MHz}$ detuned from the center frequency  of the detected photon pulse $\omega_{\rm r}/2\pi$. Note that we have operated the parametric amplifier with a relatively moderate gain of $G(\omega_{\rm r})=16.5 \,{\rm dB}$ at the center frequency of the radiation field to be detected, in order to have a flat gain curve over the entire band of detection. In this regime, the effective detection efficiency  is an order of magnitude higher than for typical setups using transistor based amplifiers only \cite{Eichler2011}. After amplification, we record the time dependent quadrature amplitudes $\{X(t), P(t)\}$ in a microwave frequency heterodyne detection setup similar to the ones discussed in Ref.~\cite{Wallraff2004}. Performing temporal mode matching by convolving $\{X(t), P(t)\}$ with an appropriate filter function \cite{Eichler2012}, we retain one pair of values $\{X,P\}$ per generated Bell state.

After the field detection we perform qubit state tomography (Fig.~\ref{fig:setup}(d)). We measure the qubit Bloch vector components $\langle\sigma_x\rangle$, $\langle\sigma_y\rangle$ and $\langle\sigma_z\rangle$ by rotating the qubit into the respective eigenbasis and then applying a coherent read-out tone to the resonator. \CE{Due to the dispersive resonator frequency shift of $\chi/2\pi =2.1 \,{\rm MHz}$ the integrated phase quadrature $Q$ of the transmitted time-dependent signal $Q(t)$ depends on the measured qubit state~\cite{Vijay2011,Wallraff2005,Gambetta2007}. The  probability distribution $p(Q)$ is fitted to a weighted sum of two independently measured reference distributions for the ground and excited state to extract the excited state population in the chosen basis (Fig.~\ref{fig:setup}(e)).}
\AW{Due to qubit decay during the time required for the measurement of photon field quadratures after preparation of the entangled state, the single-shot qubit readout fidelity making use of the same mode is limited to $37\%$. In future experiments this aspect could be improved by using separate modes for photon generation and qubit read-out, similar to Ref.~\cite{Leek2010}.}

We extract \AW{the} correlations between qubit and photon in the generated Bell states by recording 3-dimensional histograms of triplets $\{X,P,Q\}$, which count the number of times for which the qubit read-out quadrature $Q$ is measured in combination with photon field quadratures $X$ and $P$.  \AW{Within the limitations of the available memory, we chose to discretize the histograms into $128\times128$ bins for the measured photon field quadratures times $8$ bins for the measured qubit quadrature.} \CE{Efficient data acquisition \AW{and generation of histograms is realized in real-time} using field programmable gate array electronics~\cite{Bozyigit2011}.} The resulting data, which is obtained  after preparation and detection of $\sim 3 \times 10^8$ Bell states, contains complete information about the photon statistics as well as all relevant qubit-photon correlations. From the measured histograms we extract the qubit population for each quadrature pair \{$X$,$P$\} \CE{by fitting the histogram columns along the $Q$ \AW{axis} to the two reference histograms shown in Fig.~\ref{fig:setup}(e)}.

\begin{figure*}[t]
\centering
\includegraphics[scale=1]{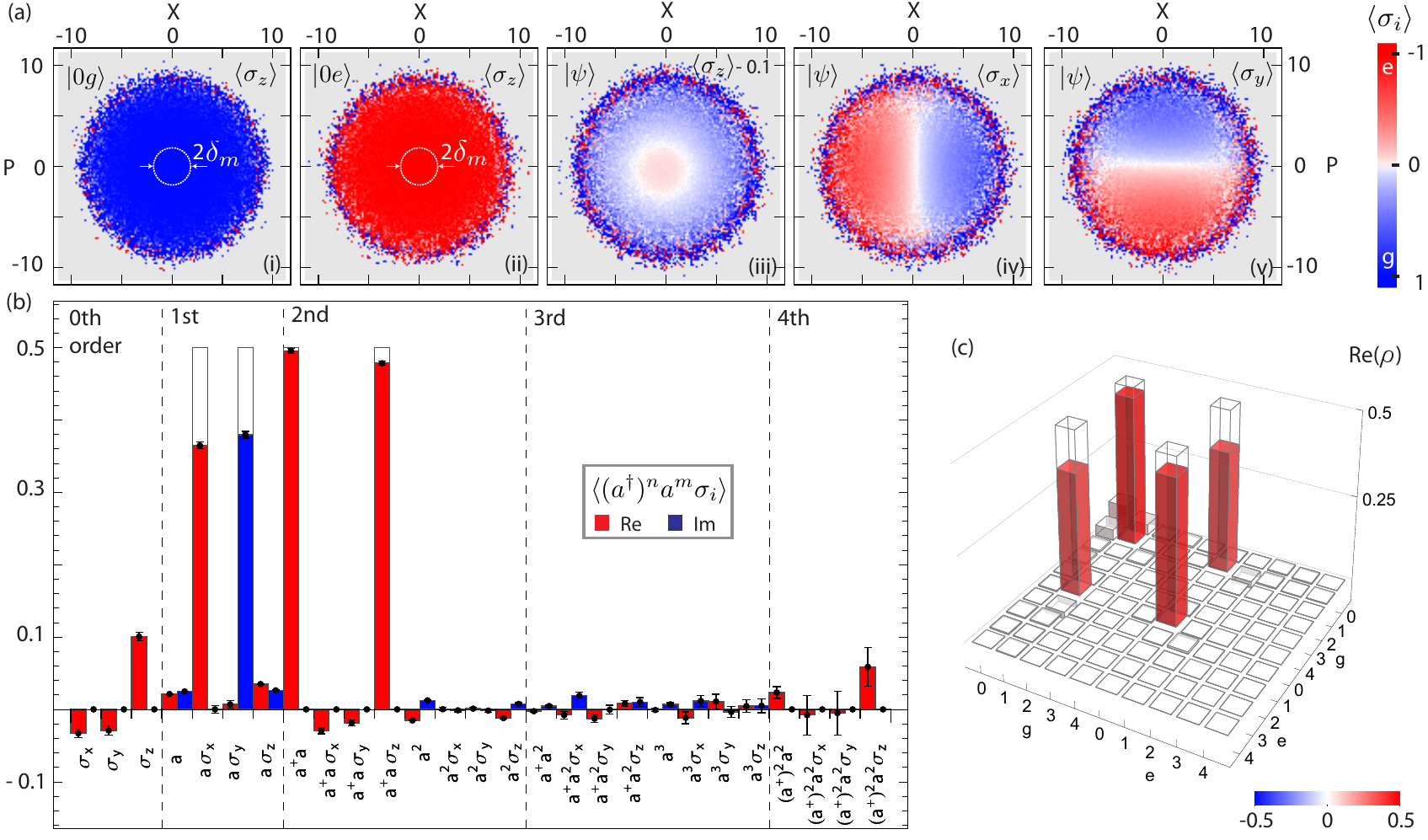}
\caption{{Photon-qubit correlations for a prepared Bell state}. ({a}) Qubit state population conditioned on the measured photon field quadratures $X$ and $P$ for the indicated Bloch vector components.  $\{X,P\}$ pairs for which no measurement results occured are shown in gray. White circles indicate the standard deviation of the photon field distribution $\delta_m$. (i)-(iii) $\langle\sigma_z\rangle$ for the reference states $|0g\rangle$, $|0e\rangle$ and the Bell state $|\psi\rangle$. \AW{For better visibility only, the data in subpanel (iii) is offset by its total mean~$\approx0.1$ compensating for the qubit decay during photon detection.} (iv)-(v) $\langle \sigma_x \rangle$ and $\langle \sigma_y \rangle$ for the Bell state $|\psi \rangle$.  ({b}) Expectation values $\langle (a^\dagger)^n a^m \sigma_i \rangle $ extracted from the qubit-photon field correlations shown in {A} and the measured photon field distribution. The real (imaginary) part of these measured moments are shown in red (blue) and compared to the ideal Bell state (wireframe). The error bars are extracted from the standard deviation of repeated measurements. ({c}) Real part of the measured (solid) and ideal (wireframe) density matrix $\rho$ for the Bell state $|\psi\rangle$ \AW{with fidelity $F=\langle\psi|\rho|\psi\rangle = 83\%$}.}
\label{fig:Bell state summary}
\end{figure*}

Preparing two reference states $|0g\rangle$ and $|0e\rangle$ -- for which the photon field is left in the vacuum state and thus is not correlated with the qubit -- we find that the qubit population (blue: ground state, red: excited state) is independent of the detected field quadratures $X$ and $P$ (see Fig.~\ref{fig:Bell state summary}(a)~(i)-(ii)). This also indicates that there are no correlations of technical origin between the qubit and photon field measurements.  The standard deviation  of the photon field distribution  \CE{$\delta_m = 1.84 $} \AW{is larger than the quantum limit $1/\sqrt{2}$, due to  noise added by the amplifiers, losses in the cables and microwave components, as well as finite mode matching efficiency, the combination of which corresponds to an effective detection efficiency of $\eta=15\%$.} Individual measurement results with large amplitude values ($\sqrt{X^2 + P^2} \gtrsim 3\delta_m $) are unlikely, which causes the larger statistical uncertainty in the extracted qubit populations at the boundary of the colored regions.

When Bell states $|\psi\rangle=(|0e\rangle+|1g\rangle)/\sqrt{2}$ are prepared we find a clear dependence of the measured qubit Bloch vector on the measured field quadratures \{$X$, $P$\} (Fig.~\ref{fig:Bell state summary}(a)~(iii)-(v)), in stark contrast to the results obtained for separable states. Measuring the qubit in the $\sigma_z$ basis we find a higher probability to observe the qubit in its ground state at large measured field amplitudes (blue region) and a higher probability to find the qubit in the excited state at small measured field amplitudes (red region in Fig.~\ref{fig:Bell state summary}(a)~(iii)). This observation is consistent with the expectation to either find the qubit in the ground state when a photon is propagating in mode $a$ or in the excited state when $a$ is in the vacuum state resulting in correspondingly small field amplitudes. The fact that the measured qubit population is circularly symmetric in phase space, i.e.~it is independent of the phase of the propagating field, indicates that $|g\rangle$ and $|e\rangle$ are correlated with Fock states -- such as the single photon or the vacuum state.

To distinguish the coherent superposition of $|0e\rangle$ and $|1g\rangle$ in the Bell state from a mere statistical mixture, we measure the equatorial components $\langle\sigma_x\rangle$ and $\langle\sigma_y\rangle$ of the qubit Bloch vector by applying $\pi/2$ pulses to the qubit about the corresponding axes and determine their correlations with the measured radiation field. We find that whenever a positive  field quadrature $X > 0$ is measured the qubit is more likely to be found in a state with positive $\langle \sigma_x\rangle$ (blue region) and vice versa for negative values (red region in Fig.~\ref{fig:Bell state summary}(a)~(iv)). This observation can be understood, when rewriting the Bell state $|\psi\rangle = [(|0\rangle + |1\rangle)|g_x\rangle + (|1\rangle - |0\rangle)|e_x\rangle]/2$ in the eigenbasis $\{|g_x\rangle,|e_x\rangle \}$ of the measurement observable $\sigma_x$: We note that for the field component $(|0\rangle + |1\rangle)$ we find $\langle{X}\rangle > 0$ which is correlated with the state $|g_x\rangle$, while $|e_x\rangle$ is correlated with $(|0\rangle - |1\rangle)$ for which $\langle{X}\rangle < 0$. Equivalently, the $\langle \sigma_y \rangle$ component is correlated with the sign of the $P$ quadrature measurement as shown in Fig.~\ref{fig:Bell state summary}(a)~(v).

Already in the raw measurement data \AW{we clearly observe} the expected qubit-photon correlations including their phase coherence. In order to further quantify the properties of the prepared Bell state we evaluate the statistical moments $\langle(a^\dagger)^n a^m \sigma_i\rangle$ from the measured set of 3-dimensional histograms \CE{using the methods presented in Refs.~\cite{Eichler2012,Eichler2011}}. The resulting measured expectation values (colored bars) of products between the Pauli operators $\sigma_i$ and photon field operators $a$, $a^\dagger$  are compared with the theoretical values of an ideal Bell state (wireframes) up to order \CE{$n+m =4$} in Fig.~\ref{fig:Bell state summary}(b). Here we note that in comparison to earlier measurements \cite{Bozyigit2011,Eichler2011}, the increase in detection efficiency enabled by the parametric amplifier is essential for the measurement of higher order expectation values \AW{which now also include} products of qubit and photon field operators.

The measured zeroth order moments $\langle\sigma_i\rangle$ represent the Bloch vector of the qubit. Since all values are close to zero the qubit is, as expected, in the maximally mixed state when the photon part of $|\psi\rangle$ is traced out. \CE{The small finite value of $\langle\sigma_z\rangle$ \AW{is due to} qubit decay during the time between state preparation and qubit tomography \AW{required for performing photon tomography in the same mode}}. Finite first order expectation values $\langle a \sigma_i \rangle$ are due to the \AW{expected} correlations between the equatorial component of the Bloch vector and the phase of the photon field, already observed in the raw measurement data (Fig.~\ref{fig:Bell state summary}(a) (iv)-(v)). The finite second order moments \AW{show that} the single excitation is shared among qubit and photon field. Since the mean product of excitations $\langle a^\dagger a \sigma_z \rangle$ is close to the mean photon number $\langle a^\dagger a\rangle$ we find that whenever a photon is detected, the qubit is in the ground state for which $\sigma_z$ takes the value 1.  We also find that all higher order moments with $n+m =3,4$ are close to zero within their statistical errors, indicating that the photon field is a superposition of vacuum and single-photon states only \cite{Eichler2012}. In particular the measured anti-bunching ($\langle (a^\dagger)^2 a^2\rangle= 0.023\pm0.008$) shows that there are no contributions of higher photon number states \cite{Bozyigit2011}. Moments of higher order can also be determined from the measured histogram data \CE{(not shown)}, albeit with statistical errors which depend exponentially on increasing order~\cite{daSilva2010}.

\begin{figure}[!t]
\centering
\includegraphics[scale=1]{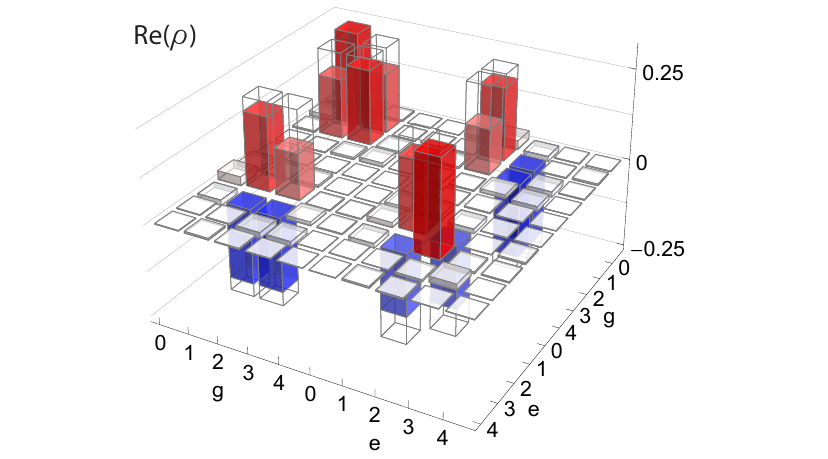}
\caption{{ Density matrix for a two-photon entangled state.} Real part of the measured (solid) and the ideal (wireframe) density matrix for the state $|\phi\rangle = \frac{1}{2}(|1\rangle+|2\rangle)|g\rangle +\frac{1}{2}(|1\rangle-|2\rangle)|e\rangle$ generated and measured with fidelity $F=80\%$. }
\label{fig:TwoPhotonState}
\end{figure}

We have also evaluated the density matrix $\rho$ of the joint qubit-photon state from the measurement data using a direct linear mapping from the moments to the density matrix elements~\cite{Eichler2012}. In order to make use of the full measurement data and to guarantee a completely positive density matrix we additionally apply a maximum-likelihood procedure which estimates the most likely density matrix from measured moments and their respective standard deviations up to order $n+m=8$. This allows for reconstructing the density matrix in a 10-dimensional Hilbert space including photon number states up to $|n\rangle=|4\rangle$. As already expected from the vanishing fourth order moments, we find all number state populations with $n>1$ close to zero (Fig.~\ref{fig:Bell state summary}(c)).  The coherent superposition of the two contributing basis states $|0e\rangle$ and $|1g\rangle$ is reflected in the large off-diagonal elements. Adjusting the overall local oscillator phase, the elements of the imaginary part of the density matrix (not shown) have been minimized to less than $0.023$. The total fidelity of the reconstructed state compared to the ideal Bell state $|\psi\rangle$ is $F=\langle\psi|\rho|\psi\rangle = 83\%$. The loss of fidelity is dominantly due to qubit decay and decoherence during the $60\,{\rm ns}$ period between the state preparation pulse and the final tomography pulse, which determines the time at which the qubit state is characterized. As a measure of entanglement between the photon states $\{|0\rangle,|1\rangle\}$ and the qubit states $\{|g\rangle,|e\rangle\}$ we determine the concurrence $C(\rho) = 0.72$ which is clearly above the entanglement threshold $C(\rho)=0$.

\AW{To demonstrate the versatility of our state preparation, detection and reconstruction scheme beyond existing experiments we have prepared entangled states between stationary qubits and multiple propagating photons such as $|\phi\rangle = \frac{1}{2}(|1\rangle+|2\rangle)|g\rangle +\frac{1}{2}(|1\rangle-|2\rangle)|e\rangle$. Making use of the third energy level $f$ of the transmon~\cite{Bianchetti2010},} we start the preparation sequence with a two-photon $\pi$-pulse on the $g$-$f$ transition which transforms the state from $|0g\rangle \rightarrow |0f\rangle$. We then apply a sequence of two flux pulses to the transmon.  Tuning the $e$-$f$ transition in resonance with $\omega_{\rm r}/2\pi$ during the first pulse, one excitation is swapped into the resonator $|0f\rangle\rightarrow|1e\rangle$.  The second pulse entangles the state $|1e\rangle$ with $|2g\rangle$ by tuning the $g$-$e$ transition in resonance with the cavity for an appropriate time. An additional $\pi/2$ pulse applied at the $g$-$e$ transition frequency creates the state $|\phi\rangle$. Note that this state preparation sequence can be interpreted as the generation of the separable state $\frac{1}{\sqrt{2}}(|1\rangle+|2\rangle)\otimes(|g\rangle +|e\rangle)$ and an entangling controlled phase gate, which changes the sign of the two photon component $|2\rangle$ only if the qubit is in the excited state. The entanglement between photon field and qubit thus becomes apparent in the negative sign of the $|2 e\rangle$ component in $|\phi\rangle$. We characterize the prepared state using the  methods described above, which results in a final density matrix with fidelity $F=80\%$ compared to the ideal one (Fig.~\ref{fig:TwoPhotonState}). \AW{The sign change apparent in the reconstructed state as the six negative (blue) elements clearly demonstrates the entanglement of the propagating multi-photon state with the stationary qubit.}

\CE{In our experiments, we have demonstrated the generation and detection of entanglement between a superconducting qubit and a propagating microwave field. The development of sensitive detection techniques for the measurement of  photon-qubit quantum correlations is an important step towards using itinerant microwave photons as a quantum information carrier, e.g.~for connecting spatially separated superconducting circuits or other systems interacting with microwave photons. The development of microwave photon counters~\cite{Chen2011a} and the high level of control achievable over superconducting circuits put the realization of microwave photon based quantum network experiments within reach.}
\\\\
Acknowledgments: We are grateful for discussions with Denis Vion, Marcus da Silva and Stephan Ritter. This work was supported by the European Research Council (ERC) through a Starting Grant and by ETHZ.

%
\bibliographystyle{science}

\end{document}